\documentstyle[epsfig,preprint,aps]{revtex}  
\begin{document} 
\tighten      

\def\bea{\begin{eqnarray}}
\def\eea{\end{eqnarray}}
\def\beas{\begin{eqnarray*}}
\def\eeas{\end{eqnarray*}}
\def\nn{\nonumber}
\def\ni{\noindent}
\def\G{\Gamma}
\def\d{\delta}
\def\l{\lambda}
\def\L{\Lambda}
\def\g{\gamma}
\def\m{\mu}
\def\n{\nu}
\def\s{\sigma}
\def\tt{\theta}
\def\b{\beta}
\def\a{\alpha}
\def\f{\phi}
\def\fh{\phi}
\def\y{\psi}
\def\z{\zeta}
\def\p{\pi}
\def\e{\epsilon}
\def\ve{\varepsilon}
\def\cl{{\cal L}}
\def\cv{{\cal V}}
\def\cz{{\cal Z}}
\def\co{{\cal O}}
\def\pl{\partial}
\def\ov{\over}
\def\~{\tilde}
\def\rar{\rightarrow}
\def\lar{\leftarrow}
\def\lrar{\leftrightarrow}
\def\rra{\longrightarrow}
\def\lla{\longleftarrow}
\def\8{\infty}
\def\h{\hbar}
\def\lg{\langle}
\def\rg{\rangle}
\def\ag{\biggl({m_\f^2 \ov \m_*^2}\biggr)}
\def\agg{\biggl({m_\f^2 \ov \m^2}\biggr)}

\def\prd{Phys. Rev. D}
\def\prl{Phys. Rev. Lett.} 
\def\np{Nucl. Phys.}
\def\plb{Phys. Lett. B}
\def\ap{Ann. Phys. (N. Y.)} 
\def\ed{\end{document}}

\title{RG-Improved Three-Loop Effective Potential of the Massive 
$\boldmath\phi^4\unboldmath$ Theory}

\author{J.-M. Chung\footnote{Email address: jmchung@khu.ac.kr}}
\address{Research Institute for Basic Sciences
and Department of Physics,\\ Kyung Hee University, Seoul 130-701, Korea}
\author{B. K. Chung\footnote{Email address: bkchung@khu.ac.kr}}
\address{Asia Pacific Center for Theoretical Physics, Seoul 135-080, 
Korea,\\
and Research Institute for Basic Sciences
and Department of Physics,\\ Kyung Hee University, Seoul 130-701,
Korea}

\maketitle
\vspace{2cm}

\begin{abstract}  
\indent The renormalization group method is applied to the
three-loop effective potential of the massive $\phi^4$ theory in
the $\overline{\rm MS}$ scheme in order to obtain the
next-next-next-to-leading logarithm resummation. For this, we
exploit four-loop parts of the renormalization group functions
$\beta_\lambda$, $\gamma_m$, $\gamma_\phi$, and $\beta_\Lambda$, 
which were already given to
five-loop order via the renormalization of the zero-, two-, and
four-point one-particle-irreducible Green's functions, to solve
evolution equations for the parameters $\lambda$, $m^2$, $\phi$, and
$\Lambda$ within the accuracy of the three-loop order.
\end{abstract}


%%%%%%%%%%%%%%%%%%%%%%%%%%%%%%%%%%%%%%%%%%%%%%%%%%%%%%%%%%%%%%%%%%%%%%%%%%%%%
\section{Introduction}
%%%%%%%%%%%%%%%%%%%%%%%%%%%%%%%%%%%%%%%%%%%%%%%%%%%%%%%%%%%%%%%%%%%%%%%%%%%%%

The renormalization group (RG) method has proved one of the most
important tools in refined perturbative analysis. The concept of
the RG-improved perturbation theory was originally introduced
long ago within the context of quantum electrodynamics (QED)
in the landmark work of
Gell-Mann and Low \cite{gl}. In the expression of an RG-improved
quantity, whether it be the Green's function, the effective
potential, or any other quantity predictable from Feynman diagram
perturbation theory, the bare parameters in the corresponding
expression are replaced with their scale-dependent running forms
which are
usually calculated to some given order in the perturbation theory.

In one of the early applications of the RG method, Coleman
and Weinberg\cite{cw} considered the effective potential $V(\f)$
for a spacetime-independent scalar field $\f$ in the context of 
massless models. In the massive case, it has been demonstrated that 
this treatment also
works provided one takes into account a nontrivial running of the vacuum 
energy \cite{ks1,bkmn,fjse,bkmn2}.\footnote{
While in flat spacetime this running of vacuum
energy is more a tool of calculational convenience, in curved spacetime it
describes the running of the cosmological constant \cite{eo}.}

In this paper, we extend the earlier work of the present authors \cite{cc3}, 
in which the next-next-to-leading logarithm resummation of the effective
potential for a single-component massive $\f^4$ theory was
obtained, to the next-next-next-to-leading logarithm order. In
Sec.~II, without discussing the technical details of how the
effective potential is computed, a summary of the $\overline{\rm
MS}$ three-loop effective potential is given. In Sec.~III, the
perturbation solutions for the running parameters $\bar{\l}(t)$,
$\bar{m}^2(t)$, $\bar{\f}(t)$, and $\bar{\L}(t)$ are obtained, and
the result for the next-next-next-to-leading logarithm resummation is
reported. The final section is devoted to concluding remarks. 
In the appendix, we quote the previous result given in Ref.~8 
for the next-next-to-leading logarithm resummation of the effective 
potential.

%%%%%%%%%%%%%%%%%%%%%%%%%%%%%%%%%%%%%%%%%%%%%%%%%%%%%%%%%%%%%%%%%%%%%%%%%%%%
\section{Three-Loop Effective Potential in the 
$\boldmath\overline{\rm MS}\unboldmath$ Scheme}
%%%%%%%%%%%%%%%%%%%%%%%%%%%%%%%%%%%%%%%%%%%%%%%%%%%%%%%%%%%%%%%%%%%%%%%%%%%%
Let us consider a (single-component) massive $\f^4$ theory defined
by the following Lagrangian:
 \bea
 \cl&=&{1\ov 2}(\pl \f)^2-{m^2\ov 2} \f^2-{\m^{4-n}\l\ov 4!}\f^4-\L\nn\\
 &&+{\d Z\ov 2}(\pl \f)^2-{\d m^2\ov 2} \f^2
 -{\m^{4-n}\d\l\ov 4!}\f^4-\d\L\;.  \label{lg}
 \eea
Here, $\d Z$, $\d m^2$, $\d \l$, and $\d \L$ are the so-called
counterterms of the wave function, mass,  coupling constant, and
vacuum energy density, respectively. The three-loop effective
potential of this theory was calculated \cite{cc1} in the
framework of the dimensional regularization \cite{tv}, in which
an arbitrary constant, $\m$, with mass dimension is introduced
inevitably for a dimensional reason. The subtraction done in
Ref.~9 is nonminimal. This means that various
counterterms of each loop order contain mass-dependent arbitrary
finite terms, as well as $\ve$-pole terms. These finite parts of
counterterms are determined by imposing the renormalization
conditions on the effective potential at a given renormalization
scale. We stress that the dimensional regularization is perfectly
possible {\em with} renormalization conditions; the renormalized
quantities, such as effective potentials or Green's functions,
would then be identically the same as those found by
regularization with a cutoff: they depend only on the
renormalization conditions, and not on the regularization
procedure.

The calculations of effective potential in Ref.~9 and in
Refs.~11 and 12 are done in the dimensional regularization
scheme, with a specific set of renormalization conditions. The
same calculations at a lower-loop level, in the cutoff
regularization, with the same renormalization conditions can be
found in Ref.~13 and Ref.~14, respectively. We see
that the results agree with each other. Therefore, in the
mass-dependent scheme, we do not need to calculate finite parts of
three-loop diagrams. Knowledge of pole terms is sufficient.

However, in a mass-independent scheme, such as the MS or
$\overline{\rm MS}$ scheme, we have to calculate three-loop
diagrams to the $\ve^0$ order. Without imposing renormalization
conditions at a specific scale, we just leave $\m$ unspecified, as
in Eq.~(\ref{v0123}) below. This has the drawback that it does
not involve true physical parameters measured at a given scale.
Though it normally takes some effort to express physically
measurable quantities in terms of the parameters in the
mass-independent scheme, the RG equation in this scheme is dealt 
with much easier and the calculations in complicated theories are 
much more convenient.

Finite parts, as well as $\ve$-pole parts, of all genuine three-loop
integrals -- genuine in the sense that they cannot be factorized
into lower-loop integrals -- needed for the computation of the
$\overline{\rm MS}$ effective potential of the massive $\f^4$ 
theory were calculated
{\em analytically} recently \cite{jkps2}. Once all values of the diagrams
needed for the three-loop effective potential are known, the
renormalization of the three-loop effective potential is
straightforward, albeit long. Thus, we simply summarize the
renormalized result:
 \bea
 V&=&V^{(0)}+\h V^{(1)}+\h^2 V^{(2)}+\h^3 V^{(3)}+O(\h^4)\;,\nn\\
 V^{(0)}&=&{m^2 \f^2\ov 2}+{\l \f^4\ov 4!}+\L\;,\nn\\
 V^{(1)}&=&{\l\ov (4\p)^2}\biggl[-{3m^4\ov 8\l}-{3 m^2 \f^2\ov 8}
 -{3\l\f^4\ov 32}
 +\biggl\{{m^4\ov 4\l}+{m^2 \f^2\ov 4}
 +{\l\f^4\ov 16}\biggr\}\ln\agg\biggr]\;,\nn\\
 V^{(2)}&=&{\l^2\ov (4\p)^4}\biggl[{m^4\ov 8\l}
 +m^2\f^2\biggl( {3\ov 4}-{1\ov 2\sqrt{3}}
 {\rm Cl}_2\Bigl({\p\ov 3}\Bigr)\biggr)
 +\l\f^4\biggl({11\ov 32}-{1\ov 4\sqrt{3}}
 {\rm Cl}_2\Bigl({\p\ov 3}\Bigr)\biggr)\nn\\
 &&-\biggl\{{m^4\ov 4\l}+{3m^2\f^2\ov 4}
 +{5\l\f^4\ov 16}\biggr\}\ln\agg
 +\biggl\{{m^4\ov 8\l}+
 {m^2 \f^2\ov 4}+{3\l\f^4\ov 32}\biggr\}\ln^2\agg\biggr]\;,\nn\\
 V^{(3)}&=&{\l^3\ov (4\p)^6} \biggl[{m^4\ov 576\l}
 + m^2\f^2\biggl(-{2363\ov 576}
 +{13\ov 4\sqrt{3}}{\rm Cl}_2\Bigl({\p\ov 3}\Bigr)
 +{3\z(3)\ov 4}\biggr)\nn\\
 && +\l\f^4\biggl(-{4487\ov 2304}
 +{11\ov 8\sqrt{3}}{\rm Cl}_2\Bigl({\p\ov 3}\Bigr)
 +{1\ov 6}{\rm Cl}_2^2\Bigl({\p\ov 3}\Bigr)
 -{2\ov 3}{\rm Li}_4\Bigl({1\ov 2}\Bigr) +{17\z(4)\ov 24}\nn\\
 &&+{\p^2\ln^2 2\ov 36}-{\ln^4 2\ov 36}\biggr)
 +\biggl\{{41m^4\ov 96\l}
 +m^2\f^2\biggl( {371\ov 96}-{7\ov 4\sqrt{3}}
 {\rm Cl}_2\Bigl({\p\ov 3}\Bigr)\biggr)\nn\\
 &&+\l\f^4\biggl({701\ov 384}
 -{3\sqrt{3}\ov 4}{\rm Cl}_2\Bigl({\p\ov 3}\Bigr)
 +{\z(3)\ov 4}\biggr)\biggr\}\ln\agg
 -\biggl\{{17m^4\ov 48\l}+{37m^2\f^2\ov 24}\nn\\
 &&+{143\l\f^4\ov 192}\biggr\}\ln^2\agg
 +\biggl\{{5m^4\ov 48\l}+{7m^2\f^2\ov 24}
 +{9\l\f^4\ov 64}\biggr\}\ln^3\agg\biggr]\;, \label{v0123}
 \eea
where $m_\f^2$ is defined as $ m_\f^2\equiv m^2+\l\f^2/2$, 
and the two transcendental numbers ${\rm Cl}_2(\p/3)$ and 
${\rm Li}_4(1/2)$ are the Clausen's dilogarithm
and the generalized log-sine integral respectively, 
whose numerical values  are ${\rm Cl}_2(\p/3)=1.014\,941\,606...$
and ${\rm Li}_4(1/2)=0.517\,479\,061...$

%%%%%%%%%%%%%%%%%%%%%%%%%%%%%%%%%%%%%%%%%%%%%%%%%%%%%%%%%%%%%%%%%%%%%%%%%%%%
\section{Next-Next-Next-To-Leading Logarithm Resummation of the
Effective Potential}
%%%%%%%%%%%%%%%%%%%%%%%%%%%%%%%%%%%%%%%%%%%%%%%%%%%%%%%%%%%%%%%%%%%%%%%%%%%%
In the usual loop expansion, the $l$-loop quantum correction to
the effective potential for a single-component massive $\f^4$
theory has the following structure \cite{ks1}:
 \bea
 V^{(l)}(\f,\l,x,y)=\l^{l+1}\f^4\sum_{m=0}^{l-1}\sum_{n=0}^l
 a_{lmn}x^{m-2}y^n\;, \label{vl}
 \eea
where
 \bea
 x\equiv {1\ov 1+2m^2/(\l \f^2)}\;, ~~~y\equiv \ln{m_\f^2\ov \m^2}\;.
 \label{vlxy}
 \eea
With this observation, we can rearrange the order of summation
over indices $\{l,m,n\}$ appearing in the expansion of the full
effective potential, $V=\sum_{l=0}^\8 \h^l V^{(l)}$, so as to
give a leading-logarithm expansion [3, 4, 8]:
 \beas
  V=\sum_{l=0}^\8 \h^l V^{(\!(l)\!)}\;,
 \eeas
where the $l$th-to-leading logarithm contribution,
$V^{(\!(l)\!)}$, is given as follows:
 \beas
 V^{(\!(l)\!)}=\l\f^4\sum_{n=l}^\8\l^n\h^{n-l} y^{n-l}
 \sum_{m=0}^{n-1}a_{nm(n-l)}x^{m-2}\;. 
 \eeas
The concept of the leading-logarithm expansion can be explained
best by the diagram given in Fig.~1. The coefficients $G_n^{(l)}$
of $y$ in Fig.~1 are defined as follows:
\beas
G_n^{(l)}=(4\p)^{2l}\l\f^4\sum_{m=0}^{l-1}a_{lm(l-n)}x^{m-2}\;.
\eeas

The zeroth-to-leading (i.e., leading) logarithm term, $V^{(\!(0)\!)}$, 
and the first-to-leading (i.e., next-to-leading) logarithm term, $V^{(\!(1)\!)}$, 
were obtained in Ref.~3. The second-to-leading logarithm 
(i.e., next-next-to leading) logarithm term, $V^{(\!(2)\!)}$, 
was obtained in Ref.~8. Now in the present paper, we
calculate the third-to-leading, i.e., next-next-next-to-leading,
logarithm term, $V^{(\!(3)\!)}$. In order to obtain a
renormalization-group-improved effective potential which is exact
up to $L$th-to-leading logarithm order, we need $(L+1)$-loop RG
functions together with the $L$-loop effective potential \cite{bkmn}.
The various $\b$ and $\g$ functions ($\b_{\l}$,
$\g_m$, $\g_\f$, and $\b_\L$) are known up to the five-loop order
through the renormalization of the zero-, two-, and four-point
one-particle-irreducible Green's functions $\G^{(0)}$, $\G^{(2)}$,
and $\G^{(4)}$, for a massive $O(N)$ $\f^4$ theory in four dimensions
of spacetime \cite{sf,sft,ks2}. For $N=1$, their values are given 
as follows:
 \bea
 \b_\l&=&{3\l^2\h\ov(4\p)^2}-{17\l^3\h^2\ov 3(4\p)^4}+{\l^4\h^3\ov
 (4\p)^6}\biggl({145\ov 8}+12\z(3)\biggr)\nn\\
 &\equiv&\b_1\l^2\h+\b_2\l^3\h^2
 +\b_3\l^4\h^3+\b_4\l^5\h^4\;,\nn\\
 \g_m&=&{\l\h\ov (4\p)^2}-{5\l^2\h^2\ov 6(4\p)^4}
 +{7\l^3\h^3\ov 2(4\p)^6}-{\l^4\h^4\ov (4\p)^8}\biggl({477\ov 32}
 +{3\z(3)\ov 2}+3\z(4)\biggr)\nn\\
 &\equiv&\g_{m1}\l\h+\g_{m2}\l^2\h^2
 +\g_{m3}\l^3\h^3+\g_{m4}\l^4\h^4\;,\nn\\
 \g_\f&=&{\l\h\ov (4\p)^2}\times 0+{\l^2\h^2\ov 12(4\p)^4}
 -{\l^3\h^3\ov 16(4\p)^6}+{65\l^4\h^4\ov 192(4\p)^8}\nn\\
  &\equiv&\g_1\l\h+\g_2\l^2\h^2
 +\g_3\l^3\h^3+\g_4\l^4\h^4\;,\nn\\
 \b_\L&=&{m^4\h\ov 2(4\p)^2}+{m^4\l\h^2\ov (4\p)^4}\times 0
 +{m^4\l^2\h^3\ov 16(4\p)^6}
 +{m^4\l^3\h^4\ov (4\p)^8}\biggl({\z(3)\ov 2}
 -{25\ov 24}\biggr)\nn\\
  &\equiv&m^4(\b_{\L1}\h+\b_{\L2}\l\h^2
 +\b_{\L3}\l^2\h^3+\b_{\L4}\l^3\h^4)\;.
 \label{rgfunc}
 \eea\\

%%%%%%%%%%%%%%%%%%%%%%%%%%%%%%%%%%%%%%%%%%%%%%%%%%%%%%%%%%%%%%%%%%%%%%%%%%%%
{\bf 1. Running Parameters Perturbed up to Three-Loop Order}\\
%%%%%%%%%%%%%%%%%%%%%%%%%%%%%%%%%%%%%%%%%%%%%%%%%%%%%%%%%%%%%%%%%%%%%%%%%%%%

Since we assume the effective potential $V(=\sum_{l=0}^\8 \h^l V^{(l)}=
\sum_{l=0}^\8\h^l V^{(\!(l)\!)}$) is independent of the
renormalization scale $\m$ for the fixed values of the bare parameters,
arbitrary changes of this scale $\m$ can be compensated for by
appropriate (finite) changes in the quantities ($\l$, $m^2$, $\f$, and $\L$)
that characterize the theory. This leads to
the RG equation for the effective potential $V(\m,\l,m^2,\f,\L)$:
 \bea
 \biggl[\m{\pl\ov \pl\m}+\b_{\l}{\pl\ov \pl\l}+\g_m m^2{\pl\ov \pl m^2}
 -\g_\f\f{\pl\ov \pl\f}+\b_\L{\pl\ov\pl\L}\biggr]V
 (\m,\l,m^2,\f,\L)=0\;. \label{rge}
 \eea
Applying the method of characteristics to Eq.~(\ref{rge}), we can write
the solution of Eq.~(\ref{rge}), $V(\m,\l,m^2,\f,\L)$ as follows:
 \bea
 V(\m,\l,m^2,\f,\L)=
 V(\bar{\mu},\bar{\l},\bar{m}^2,\bar{\f},\bar{\L})\;,
 \label{char}
 \eea
where the barred quantities are running parameters which satisfy
the following differential equations with respect to a running scale $t$:
 \bea
 &&~~~~~~~~~~~~~~~~~~
\h{d\bar{\m}\ov dt}=\bar{\m}\;,\label{mu}\\
 &&~~\h{d\bar{\l}\ov dt}=\b_\l(\bar{\l})\;,~~~~
 \h{d\bar{m}^2\ov dt}=\g_m(\bar{\l})\bar{m}^2\;,\nn\\
 &&\h{d\bar{\f}\ov dt}=-\g_\f(\bar{\l})\bar{\f}\;,~~~~
 \h{d\bar{\L}\ov dt}=\b_\L(\bar{\l},\bar{m}^2)\;,\label{bar}
 \eea
and at the boundary point, $t=0$, their values are given as
$\bar{\m}(t=0)=\m$, $\bar{m}^2(t=0)=m^2$, $\bar{\f}(t=0)=\f$,
and $\bar{\L}(t=0)=\L$.

The $\bar{\m}$ differential equation is very simple and its
solution is given as
 \bea
 \bar{\m}^2(t)=\m^2 \exp(2t/\h)\;. \label{bmu}
 \eea
For the purpose of our leading logarithm
expansion, it is sufficient to solve four equations in
Eq.~(\ref{bar}) perturbatively.\footnote{In relation to this
point, two comments are in order: (i) since we have no knowledge
on the exact $\b$ and $\g$ functions ($\b_{\l}$, $\g_m$, $\g_\f$,
and $\b_\L$), solutions to four equations in Eq.~(\ref{bar})
cannot be the exact ones even when these four linear differential
equations with $\b_{\l}$, $\g_m$, $\g_\f$, and $\b_\L$ given in
Eq.~(\ref{rgfunc}) can be integrated exactly and (ii) although
some of these four equations with $\b_{\l}$, $\g_m$, $\g_\f$, and
$\b_\L$ of Eq.~(\ref{rgfunc}) can be solved exactly, we have to
expand the obtained solutions to the given order for the leading
logarithm expansion.} In order to solve the $\bar{\l}$ differential
equation, we try a perturbative solution by writing
 \beas
 {\bar\l}={\bar \l}^{\lg 0\rg}+\h{\bar \l}^{\lg 1\rg}
 +\h^2{\bar\l}^{\lg 2\rg}+\h^3{\bar \l}^{\lg 3\rg}+O(\h^4)\;,
 \eeas
with the boundary conditions $\bar{\l}^{\lg 0\rg}(0)=\l$ and 
${\bar \l}^{\lg 
1\rg}(0)={\bar \l}^{\lg 2\rg}(0) ={\bar \l}^{\lg 3\rg}(0)=0$.
Then, with $\b_\l$ in Eq.~(\ref{rgfunc}), the equation we want to
solve is split into four first-order linear differential
equations within the desired order:
  \beas
 {d\bar{\l}^{\lg 0\rg}\ov dt}&=&\b_1\bar{\l}^{\lg 0\rg 2}\;,\nn\\
 {d\bar{\l}^{\lg 1\rg}\ov dt}&=&2\b_1\bar{\l}^{\lg 0\rg}\bar{\l}^{\lg 1\rg}
 +\b_2\bar{\l}^{\lg 0\rg 3}\;,\nn\\
 {d\bar{\l}^{\lg 2\rg}\ov dt}&=&2\b_1\bar{\l}^{\lg 0\rg}\bar{\l}^{\lg 2\rg}
 +\b_1\bar{\l}^{\lg 1\rg 2}
 +3\b_2\bar{\l}^{\lg 0\rg 2}\bar{\l}^{\lg 1\rg}+
 \b_3\bar{\l}^{\lg 0\rg 4}\;,\nn\\
 {d\bar{\l}^{\lg 3\rg}\ov dt}&=&2\b_1\bar{\l}^{\lg 0\rg}\bar{\l}^{\lg 3\rg}
 +2\b_1\bar{\l}^{\lg 1\rg}\bar{\l}^{\lg 2\rg}
 +3\b_2\bar{\l}^{\lg 0\rg 2}\bar{\l}^{\lg 2\rg}\nn\\
 &&+3\b_2\bar{\l}^{\lg 0\rg}\bar{\l}^{\lg 1\rg 2}
 +4\b_3\bar{\l}^{\lg 0\rg 3}\bar{\l}^{\lg 1\rg}
 +\b_4\bar{\l}^{\lg 0\rg 5}\;.
 \eeas
Solutions to the $\bar{\l}^{\lg 0\rg}$, $\bar{\l}^{\lg 1\rg}$,
and $\bar{\l}^{\lg 2\rg}$ differential equations have been obtained
already in Ref.~8. The $\bar{\l}^{\lg 3\rg}$ differential equation
is readily integrated as follows:
 \bea
 \bar{\l}^{\lg 3\rg}&=&{\l^4\ov (4\p)^6}\biggl\{{1\ov T^2}
 \biggl[{95807\ov 23328}+{49\z(3)\ov 9}-3\z(4)+20\z(5)\biggr]\nn\\
 &&+{1\ov T^3}\biggl[{27251\ov 5832}+{68\z(3)\ov 9}
 -\biggl({27251\ov 2916}+{136 \z(3)\ov 9}\biggr)
 \ln T\biggr]\nn\\
 &&+{1\ov T^4}\biggl[-{204811\ov 23328}
 -13 \z(3)+3 \z(4)-20\z(5)+\biggl({121057\ov 5832}
 +{68\z(3)\ov 3}\biggr)\ln T\nn\\
 &&-{24565\ov 1458}\ln^2T +{4913\ov 729}\ln^3T\biggr]\biggr\}\;,\label{bls}
 \eea
where $T\equiv 1-3\l t/(4\p)^2$. 

Similarly, we write $\bar{m}^2$
as 
 \beas {\bar m}^2={\bar m}^{2\lg 0\rg}+\h{\bar m}^{2\lg 1\rg}
 +\h^2{\bar m}^{2\lg 2\rg}+\h^3{\bar m}^{2\lg 3\rg}+O(\h^4)
 \eeas 
and obtain, with $\g_m$ in Eq.~(\ref{rgfunc}), four split
first-order linear differential equations:
 \bea
 {d\bar{m}^{2\lg 0\rg}\ov dt}&=&\g_{m1}\bar{\l}^{\lg 0\rg}\bar{m}^{2\lg 0\rg}
 \;,\nn\\
 {d\bar{m}^{2\lg 1\rg}\ov dt}&=&\g_{m1}\bar{\l}^{\lg 0\rg}\bar{m}^{2\lg 1\rg}
 +\g_{m1}\bar{\l}^{\lg 1\rg}\bar{m}^{2\lg 0\rg}
 + \g_{m2}\bar{\l}^{\lg 0\rg 2}\bar{m}^{2\lg 0\rg}\;,\nn\\
 {d\bar{m}^{2\lg 2\rg}\ov dt}&=&\g_{m1}\bar{\l}^{\lg 0\rg}\bar{m}^{2\lg 2\rg}
 +\g_{m1}\bar{\l}^{\lg 2\rg}\bar{m}^{2\lg 0\rg}
 +\g_{m1}\bar{\l}^{\lg 1\rg}\bar{m}^{2\lg 1\rg}
 +2\g_{m2}\bar{\l}^{\lg 0\rg}\bar{\l}^{\lg 1\rg}\bar{m}^{2\lg 0\rg}\nn\\
 &&+\g_{m2}\bar{\l}^{\lg 0\rg 2}\bar{m}^{2\lg 1\rg}
 +\g_{m3}\bar{\l}^{\lg 0\rg 3}\bar{m}^{2\lg 0\rg}\;,\nn\\
 {d\bar{m}^{2\lg 3\rg}\ov dt}&=&\g_{m1}\bar{\l}^{\lg 0\rg}\bar{m}^{2\lg 3\rg}
 +\g_{m1}\bar{\l}^{\lg 1\rg}\bar{m}^{2\lg 2\rg}
 +\g_{m1}\bar{\l}^{\lg 2\rg}\bar{m}^{2\lg 1\rg}
 +\g_{m1}\bar{\l}^{\lg 3\rg}\bar{m}^{2\lg 0\rg}\nn\\
 &&+\g_{m2}\bar{\l}^{\lg 0\rg 2}\bar{m}^{2\lg 2\rg}
 +2\g_{m2}\bar{\l}^{\lg 0\rg}\bar{\l}^{\lg 1\rg}\bar{m}^{2\lg 1\rg}
 +2\g_{m2}\bar{\l}^{\lg 0\rg}\bar{\l}^{\lg 2\rg}\bar{m}^{2\lg 0\rg}\nn\\
 &&+\g_{m2}\bar{\l}^{\lg 1\rg 2}\bar{m}^{2\lg 0\rg}
 +\g_{m3}\bar{\l}^{\lg 0\rg 3}\bar{m}^{2\lg 1\rg}
 +3\g_{m3}\bar{\l}^{\lg 0\rg 2}\bar{\l}^{\lg 1\rg}\bar{m}^{2\lg
 0\rg}\nn\\
 && +\g_{m4}\bar{\l}^{\lg 0\rg 4}\bar{m}^{2\lg
 0\rg}\;.\label{mbe}
 \eea
With the $\bar{\l}$ solutions ($\bar{\l}^{\lg 0\rg}$
$\bar{\l}^{\lg 1\rg}$, $\bar{\l}^{\lg 2\rg}$, and $\bar{\l}^{\lg
3\rg}$), and the lower-order $\bar{m}^2$ solutions
($\bar{m}^{2\lg 0\rg}$, $\bar{m}^{2\lg 1\rg}$, and $\bar{m}^{2\lg
2\rg}$, which have appeared also in Ref.~8,
together with the boundary condition ${\bar m}^{2\lg 3\rg}(0)=0$,
we obtain the following solution of $\bar{m}^{2\lg 3\rg}$:
 \bea
 \bar{m}^{2\lg 3\rg}&=&{\l^3 m^2\ov (4\p)^6}\biggl\{{1\ov T^{1/3}}
 \biggl[-{245089\ov 944784}-{89\z(3)\ov 54}+\z(4)-{40\z(5)\ov 9}\biggr]\nn\\
 &&+{1\ov T^{4/3}}\biggl[{539479\ov 314928}
 +{68 \z(3)\ov 27}-\z(4)+{20\z(5)\ov 3}
 +\biggl({30379\ov 314928}+{34 \z(3)\ov 81}\biggr)\ln T\biggr]\nn\\
 &&+{1\ov T^{7/3}}\biggl[{73843\ov 629856}+{11\z(3)\ov 27}
 -\biggl({38777\ov 19683}+{272 \z(3)\ov 81}\biggr)\ln T
 -{5491\ov 19683}\ln^2T\biggr]\nn\\
 &&+{1\ov T^{10/3}}\biggl[-{2968225\ov 1889568}
 -{23\z(3)\ov 18}-{20\z(5)\ov 9}+\biggl({1284061\ov 314928}
 +{238 \z(3)\ov 81}\biggr)\ln T\nn\\
 &&-{85255\ov 39366}\ln^2T
 +{68782\ov 59049}\ln^3T\biggr]\biggr\}\;.\label{bms}
 \eea

If we notice that the $\bar{\f}$ differential equation and the
$\bar{m}^2$ differential equation in Eq.~(\ref{bar}) are of the
same structure, except for the minus sign on the right-hand side, then
we can readily write down linear differential equations for ${\bar
\f}^{\lg 0\rg}$, ${\bar \f}^{\lg 1\rg}$, ${\bar \f}^{\lg 2\rg}$,
and ${\bar \f}^{\lg 3\rg}$ in the perturbative decomposition,
 \beas
 {\bar \f}={\bar \f}^{\lg 0\rg}+\h{\bar \f}^{\lg 1\rg} +\h^2{\bar
 \f}^{\lg 2\rg}+\h^3{\bar \f}^{\lg 3\rg}+O(\h^4)\;.
 \eeas
Lower-order solutions, ${\bar \f}^{\lg 0\rg}$, ${\bar \f}^{\lg
1\rg}$, and ${\bar \f}^{\lg 2\rg}$, are found in Ref.~8.
The  ${\bar \f}^{\lg 3\rg}$ solution is obtained as follows:
 \bea
 \bar{\f}^{\lg 3\rg}&=&{\l^3\f\ov (4\p)^6}\biggl\{{95\ov 46656}
 -{\z(3)\ov 27} -{7\ov 15552 T}+{1\ov T^2}\biggl[{59\ov 864}
 +{\z(3)\ov 9}-{17\ov 11664}\ln T\biggr]\nn\\
 &&+{1\ov T^3}\biggl[-{815\ov 11664}-{2\z(3)\ov 27} +{119\ov 2916}\ln T
 -{289\ov 2916}\ln^2 T\biggr]\biggr\}\;,\label{bfs}
 \eea
which satisfies the boundary condition  ${\bar \f}^{\lg 3\rg}(0)=0$.

Finally, we try the solution to the $\bar{\L}$ differential
equation as
 \beas
 {\bar \L}={\bar \L}^{\lg 0\rg}+\h{\bar \L}^{\lg 1\rg}
 +\h^2{\bar \L}^{\lg 2\rg}+\h^3{\bar \L}^{\lg 3\rg}+O(\h^4)\;.
 \eeas
The solutions to lower-order equations
 \beas
 {d\bar{\L}^{\lg 0\rg}\ov dt}&=&\b_{\L1}\bar{m}^{2\lg 0\rg 2}\;,\nn\\
 {d\bar{\L}^{\lg 1\rg}\ov dt}&=&2\b_{\L1}\bar{m}^{2\lg 0\rg}\bar{m}^{2\lg 1\rg}
 +\b_{\L2}\bar{\l}^{\lg 0\rg}\bar{m}^{2\lg 0\rg 2}\;,\nn\\
 {d\bar{\L}^{\lg 2\rg}\ov dt}&=&\b_{\L1}\bar{m}^{2\lg 1\rg 2}
 +2\b_{\L1}\bar{m}^{2\lg 0\rg}\bar{m}^{2\lg 2\rg}
 +\b_{\L2}\bar{\l}^{\lg 1\rg}\bar{m}^{2\lg 0\rg 2}\nn\\
 &&+2\b_{\L2}\bar{\l}^{\lg 0\rg}\bar{m}^{2\lg 0\rg}\bar{m}^{2\lg 1\rg}
 +\b_{\L3}\bar{\l}^{\lg 0\rg 2}\bar{m}^{2\lg 0\rg 2}\;,
 \eeas
can be found in Ref.~8. The solution to the ${\bar \L}^{\lg 3\rg}$
differential equation
 \beas
 {d\bar{\L}^{\lg 3\rg}\ov dt}&=&2\b_{\L1}\bar{m}^{2\lg 0\rg}\bar{m}^{2\lg 3\rg}
 +2\b_{\L1}\bar{m}^{2\lg 1\rg}\bar{m}^{2\lg 2\rg}
 +\b_{\L2}\bar{\l}^{\lg 2\rg}\bar{m}^{2\lg 0\rg 2}\nn\\
 &&+\b_{\L2}\bar{\l}^{\lg 0\rg}\bar{m}^{2\lg 1\rg 2}
 +2\b_{\L2}\bar{\l}^{\lg 1\rg}\bar{m}^{2\lg 0\rg}\bar{m}^{2\lg 1\rg}
 +2\b_{\L2}\bar{\l}^{\lg 0\rg}\bar{m}^{2\lg 0\rg}\bar{m}^{2\lg 2\rg}\nn\\
 &&+2\b_{\L3}\bar{\l}^{\lg 0\rg}\bar{\l}^{\lg 1\rg}\bar{m}^{2\lg 0\rg 2}
 +2\b_{\L3}\bar{\l}^{\lg 0\rg 2}\bar{m}^{2\lg 0\rg}\bar{m}^{2\lg 1\rg}
 +\b_{\L4}\bar{\l}^{\lg 0\rg 3}\bar{m}^{2\lg 0\rg 2}
 \eeas
with the boundary condition ${\bar \L}^{\lg 3\rg}(0)=0$ is given as follows:
 \bea
 {\bar \L}^{\lg 3\rg}&=&{\l^2 m^4\ov (4\p)^6}\biggl\{
 -{709\ov 720}-{73 \z(3)\ov 20}+{3\z(4)\ov 2}-{15\z(5)\ov 2}\nn\\
 &&+T^{1/3}\biggl[{592037\ov 1889568}+{305\z(3)\ov 162}-\z(4)
 +{40\z(5)\ov 9}\biggr]\nn\\
 &&+{1\ov T^{2/3}}\biggl[{1579007\ov 1259712}+ {121 \z(3)\ov 54}
 -{\z(4)\ov 2}+{10 \z(5)\ov 3}
 +\biggl({42653\ov 314928}+{34\z(3)\ov 81}\biggr)\ln T\biggr]\nn\\
 &&-{1\ov T^{5/3}}\biggl[{807937\ov 1574640}+{76\z(3)\ov 135}
 +\biggl({106301\ov 157464}+{68 \z(3)\ov 81}\biggr)\ln T
 +{5491\ov 39366}\ln^2 T\biggr]\nn\\
 &&+{1\ov T^{8/3}}\biggl[-{260647\ov 3779136}+{29\z(3)\ov 324}
 -{5\z(5)\ov 18}+\biggl({91171\ov 157464}+{34\z(3)\ov 81}\biggr)\ln T\nn\\
 &&-{2023\ov 78732}\ln^2 T+{24565\ov 118098}\ln^3 T\biggr]
 \biggr\}\;.         \label{bLs}
 \eea\\

%%%%%%%%%%%%%%%%%%%%%%%%%%%%%%%%%%%%%%%%%%%%%%%%%%%%%%%%%%%%%%%%%%%%%%%%%%%%
{\bf 2. Summary of the Next-Next-Next-to-Leading Logarithm Resummation}\\
%%%%%%%%%%%%%%%%%%%%%%%%%%%%%%%%%%%%%%%%%%%%%%%%%%%%%%%%%%%%%%%%%%%%%%%%%%%%

The key idea of the RG improvement method is that, via a judicious choice
of $t$, one can evaluate the right-hand side of Eq.~(\ref {char})
perturbatively even if large logarithms render the left-hand side
nonperturbative [4, 5]. In Ref.~4, an ambitious
choice is made so as to remove all the logarithms of the right-hand
side of Eq.~(\ref {char}). That is, $t$ is chosen so that
 \bea
 {\bar{m}^2_\f(t)\ov \bar{\m}^2(t)}={\bar{m}^2(t)+(1/2)\bar{\l}(t)
 \bar{\f}^2(t)\ov \bar{\m}^2(t)}=1\;. \label{amb}
 \eea
Although this choice gives a simple boundary function
$[\bar{\l}/(4\p)^2]^lG_l^{(l)}(\bar{\f},\bar{\l},\bar{m}^2,\bar{\L})$
for each $l$th-to-leading logarithm series $V^{(\!(l)\!)}$, it is quite
complicated to solve Eq.~(\ref{amb}) with respect to $t$. An ingenious
method which enables us to bypass this difficulty is suggested
in Ref.~4. However, that method is still awkward to work with,
even in the next-to-leading logarithm approximation. As a less implicit
choice \cite{fjse}, we choose $t$ as
 \bea
 t={\h\ov 2}\ln \biggl({m_\f^2\ov \m^2}\biggr)\;, \label{t}
 \eea
as was done in Ref.~8. While this alternative choice
does not destroy the logarithms on the right-hand side of Eq.~(\ref {char})
(thus, gives rather complicated boundary conditions), it allows us
to sum explicitly the $l$th-to-leading logarithm series
$V^{(\!(l)\!)}$ \cite{ks1,bkmn,fjse}.

From Eqs.~(\ref{bmu}) and (\ref{t}), one finds that $\bar{\m}^2(t)$ in
Eq.~(\ref{bmu}) becomes
 \bea
 \bar{\m}^2(t)=m^2+{\l\f^2\ov 2}\;, \label{bmus}
 \eea
which is independent of $\m$.
Equations (\ref{bls}), (\ref{bms}) -- (\ref{bLs}), and (\ref{bmus}),
together with the lower-loop results in Eq.~(21) of
Ref.~8, comprise the desired perturbative solutions to
the evolution equations for the running parameters. Now, all things
necessary for an improvement of the three-loop effective
potential have been obtained. We follows the same calculation
procedure as in Ref.~8 for the correct collection of
logarithms of various powers into a given leading-logarithm
series order. The calculation is straightforward. The final
result for $V^{(\!(3)\!)}(\f,\l,m^2;t)$ in the leading-logarithm
expansion
 \bea
 V&=&V^{(\!(0)\!)}(\f,\l,m^2,\L;t)+\h V^{(\!(1)\!)}(\f,\l,m^2;t)+
 \h^2 V^{(\!(2)\!)}(\f,\l,m^2;t)\nn\\
 &&+\h^3 V^{(\!(3)\!)}(\f,\l,m^2;t)+O(\h^4) \label{llse}
 \eea
is summarized as follows:
 \bea
 V^{(\!(3)\!)}&=&{\l^3\ov (4\p)^6}\biggl[{m^4\ov\l}\biggl\{
 -{709\ov 720}-{73\z(3)\ov 20}+{3\z(4)\ov 2}-{15\z(5)\ov 2}\nn\\
 &&+T^{1/3}\biggl(
 {592037\ov 1889568}+{305\z(3)\ov 162}-\z(4)+{40\z(5)\ov 9}\biggr)
 +T^{-2/3}\biggl({687889\ov 629856}+{47\z(3)\ov 27}\nn\\
 &&-{\z(4)\ov 2}+{10\z(5)\ov 3}
 +\biggl[{42653\ov 314928}+{34\z(3)\ov 81}\biggr]\ln T
 +\biggl[{2509\ov 23328}+{\z(3)\ov 3}\biggr]\ln S\biggr)\nn\\
 &&+T^{-5/3}\biggl({639571\ov 3149280}+{59\z(3)\ov 135}
 -\biggl[{53975\ov 157464}+{68\z(3)\ov 81}\biggr]\ln T
 -{5491\ov 39366}\ln^2 T\nn\\
 &&-\biggl[{4201\ov 11664}+{2\z(3)\ov 3}\biggr]\ln S
 -{323\ov 1458}\ln T\ln S-{19\ov 216}\ln^2 S\biggr)\nn\\
 &&+T^{-8/3}\biggl(-{2350933\ov 3779136}-{133\z(3)\ov 324}-{5\z(5)\ov 18}
 +\biggl[{162775\ov 157464}+{34\z(3)\ov 81}\biggr]\ln T\nn\\
 &&-{121091\ov 157464}\ln^2 T+{24565\ov 118098}\ln^3 T
 +\biggl[{991\ov 1458}+{\z(3)\ov 3}\biggr]\ln S
 -{1207\ov 1458}\ln T\ln S\nn\\
 &&+{1445\ov 2916}\ln^2 T\ln S-{115\ov 432}\ln^2 S
 +{85\ov 216}\ln T\ln^2 S+{5\ov 48}\ln^3 S\biggr)\biggr\}\nn\\
 &&+m^2\f^2\biggl\{T^{-1/3}\biggl(-{243379\ov 1889568}
 -{91\z(3)\ov 108}+{\z(4)\ov 2}-{20\z(5)\ov 9}\biggr)
 +T^{-4/3}\biggl({484237\ov 629856}\nn\\
 &&+{103\z(3)\ov 108}
 -{\z(4)\ov 2}+{10\z(5)\ov 3}+\biggl[{16541\ov 314928}
 +{17\z(3)\ov 81}\biggr]\ln T+\biggl[{973\ov 23328}\nn\\
 &&+{\z(3)\ov 6}\biggr]\ln S\biggr)
 +T^{-7/3}\biggl({96625\ov 78732}+
 {4\ov 27\sqrt{3}}{\rm Cl}_2\Bigl({\p\ov 3}\Bigr)
 +{64\z(3)\ov 27}
 -\biggl[{54553\ov 78732}+{136\z(3)\ov 81}\biggr]\nn\\
 &&\times\ln T-{2312\ov 19683}\ln^2 T
 -\biggl[{3641\ov 5832}+{4\z(3)\ov 3}\biggr]\ln S
 -{136\ov 729}\ln T\ln S-{2\ov 27}\ln^2 S\biggr)\nn\\
 &&+T^{-10/3}\biggl(-{22560307\ov 3779136}+{335\ov 108\sqrt{3}}
 {\rm Cl}_2\Bigl({\p\ov 3}\Bigr)
 -{187\z(3)\ov 108}-{10\z(5)\ov 9}\nn\\
 &&+\biggl[{262225\ov 39366}
 -{119\ov 54\sqrt{3}}{\rm Cl}_2\Bigl({\p\ov 3}\Bigr)
 +{119\z(3)\ov 81}\biggr]\ln T
 -{65314\ov 19683}\ln^2 T+{34391\ov 59049}\ln^3 T\nn\\
 &&+\biggl[{3242\ov 729}
 -{7\ov 4\sqrt{3}}{\rm Cl}_2\Bigl({\p\ov 3}\Bigr)
 +{7\z(3)\ov 81}\biggr]\ln S
 -{12155\ov 2916}\ln T\ln S+{2023\ov 1458}\ln^2 T\ln S\nn\\
 &&-{317\ov 216}\ln^2 S
 +{119\ov 108}\ln T\ln^2 S+{7\ov 24}\ln^3 S\biggr)\biggr\}
 +\l\f^4\biggl\{T^{-1}\biggl({53\ov 93312}-{\z(3)\ov 162}\biggr)\nn\\
 &&+T^{-2}\biggl({42785\ov 279936}+{5\z(3)\ov 24}-{\z(4)\ov 8}
 +{5\z(5)\ov 6}+{85\ov 15552}\ln T+{5\ov 1152}\ln S\biggr)\nn\\
 &&+T^{-3}\biggl({204905\ov 279936}
 -{1\ov 36\sqrt{3}}{\rm Cl}_2\Bigl({\p\ov 3}\Bigr)+{121\z(3)\ov 108}
 -\biggl[{31297\ov 69984}+{17\z(3)\ov 27}\biggr]\ln T\nn\\
 &&+{289\ov 17496}\ln^2 T-\biggl[{1787\ov 5184}+{\z(3)\ov 2}\biggr]\ln S
 +{17\ov 648}\ln T\ln S+{1\ov 96}\ln^2 S\biggr)\nn\\
 &&+T^{-4}\biggl(
 -{1586039\ov 559872}+{101\ov 72\sqrt{3}}{\rm Cl}_2\Bigl({\p\ov 3}\Bigr)
 +{1\ov 6}{\rm Cl}_2^2\Bigl({\p\ov 3}\Bigr)
 -{2\ov 3}{\rm Li}_4\Bigl({1\ov 2}\Bigr)\nn\\
 &&+{17\p^4\ov 2160}+{\p^2\ov 36}\ln^2 2-{1\ov 36}\ln^4 2
 -{857\z(3)\ov 648}+{\z(4)\ov 8}-{5\z(5)\ov 6}\nn\\
 &&+\biggl[{249169\ov 69984}-{17\ov 12\sqrt{3}}{\rm Cl}_2\Bigl({\p\ov 3}\Bigr)
 +{17\z(3)\ov 18}\biggr]\ln T-{122825\ov 69984}\ln^2 T
 +{4913\ov 17496}\ln^3 T\nn\\
 &&+\biggl[{2807\ov 1296}-{9\ov 8\sqrt{3}}{\rm Cl}_2\Bigl({\p\ov 3}\Bigr)
 +{3\z(3)\ov 4}\biggr]\ln S-{731\ov 324}\ln T\ln S+{289\ov 432}\ln^2 T\ln S
 \nn\\
 &&-{145\ov 192}\ln^2 S
 +{17\ov 32}\ln T\ln^2 S+{9\ov 64}\ln^3 S\biggr)\biggr\}
 +\biggl\{{m^4\ov\l}\biggl[-{19\ov 108}T^{-2/3}+T^{-5/3}\biggl(-{2\ov 27}\nn\\
 &&+{17\ov 54}\ln T+{1\ov 4}\ln S\biggr)\biggr]
 +m^2\f^2\biggl[-{2\ov 27}T^{-4/3}+T^{-7/3}\biggl(
 -{73\ov 108}+{17\ov 27}\ln T+{1\ov 2}\ln S\biggr)\biggr]\nn\\
 &&+\l\f^4\biggl[{1\ov 144}T^{-2}+T^{-3}\biggl(-{23\ov 72}
 +{17\ov 72}\ln T+{3\ov 16}\ln S\biggr)\biggr]\biggr\}{Z\ov W}\nn\\
 &&+\biggl\{{m^4\ov \l}T^{-2/3}+m^2\f^2 T^{-4/3}
 +{\l\f^4\ov 4}T^{-2}\biggr\}\biggl({P\ov 4W}
 -{Z^2\ov 8W^2}\biggr)\biggr]\;, \label{result}
 \eea
where
 \bea
 T&\equiv& 1-{3\l t\ov (4\p)^2}\;,~~~
 W\equiv m^2T^{-1/3}+{\l\f^2\ov 2}T^{-1}\;,~~~S\equiv {W\ov m^2+\l\f^2/2}\;,
\nn\\
 Z&\equiv& m^2\biggl[-{19\ov 54}T^{-1/3}+T^{-4/3}\biggl(
 {19\ov 54}+{17\ov 27}\ln T\biggr)\biggr]\nn\\
 &&+\l\f^2\biggl[{1\ov 36}T^{-1}
 +T^{-2}\biggl(-{1\ov 36}+{17\ov 18}\ln T\biggr)\biggr] \;,\nn\\
 P&\equiv&m^2\biggl[T^{-1/3}\biggl({1787\ov 11664}+{2\z(3)\ov 3}\biggr)
 -T^{-4/3}\biggl({5531\ov 5832}+{4\z(3)\ov 3}+{323\ov 1458}\ln T\biggr)\nn\\
 &&+T^{-7/3}\biggl({9275\ov 11664}+{2\z(3)\ov 3}-{221\ov 729}\ln T
 +{578\ov 729}\ln^2 T\biggr)\biggr]\nn\\
 &&+\l\f^2\biggl[{43\ov 2592}T^{-1}
 +T^{-2}\biggl(-{535\ov 432}-2\z(3)+{17\ov 324}\ln T\biggr)\nn\\
 &&+T^{-3}\biggl({3167\ov 2592}+2\z(3)-{17\ov 9}\ln T
 +{289\ov 162}\ln^2 T\biggr)\biggr]\;. \label{twszp}
 \eea
The leading, next-to-leading, and next-next-to-leading terms
[$V^{(\!(0)\!)}$, $V^{(\!(1)\!)}$, and $V^{(\!(2)\!)}$] in Eq.~(\ref{llse})
are given in Eq.~(\ref{rslt}) of the appendix.

%%%%%%%%%%%%%%%%%%%%%%%%%%%%%%%%%%%%%%%%%%%%%%%%%%%%%%%%%%%%%%%%%%%%%%%%%
\section{Concluding Remarks}
%%%%%%%%%%%%%%%%%%%%%%%%%%%%%%%%%%%%%%%%%%%%%%%%%%%%%%%%%%%%%%%%%%%%%%%%
We have applied the RG method for a systematic resummation of the
perturbation expansion. The next-next-next-to leading logarithm
resummation, $V^{(\!(3)\!)}$, given in Eq.~(\ref{result}) is our
main result. In obtaining this, we have exploited the four-loop
parts of the RG functions $\b_\l$, $\g_m$, $\g_\f$, and $\b_\L$,
which were already given to five-loop order via the
renormalization of the zero-, two-, and four-point
one-particle-irreducible Green's functions, to solve
evolution equations for the parameters $\l$, $m^2$, $\f$, and
$\L$ within the accuracy of the three-loop order.

From the conventional three-loop expansion result, the coefficients [of
the powers of $y\equiv\ln(m_\f^2/\m^2)$] \{$G_0^{(0)}$\},
\{$G_0^{(1)}, G_1^{(1)}$\}, \{$G_0^{(2)}, G_1^{(2)},
G_2^{(2)}$\}, and \{$G_0^{(3)}, G_1^{(3)}, G_2^{(3)},
G_3^{(3)}$\} in Fig.~1 can be read off from Eq.~(\ref{v0123}). 
We have mentioned in Ref.~8 that our previous 
nonperturbative-in-$t$ results
$V^{(\!(0)\!)}$, $V^{(\!(1)\!)}$, and $V^{(\!(2)\!)}$ well
reproduce the leading logarithm series \{$G_0^{(0)}$,
$G_0^{(1)}$, $G_0^{(2)}$, $G_0^{(3)}$, $\cdots$\}, \{$G_1^{(1)}$,
$G_1^{(2)}$, $G_1^{(3)}$, $\cdots$\}, and \{$G_2^{(2)}$,
$G_2^{(3)}$, $\cdots$\}, respectively. The expansion of our new 
(nonperturbative-in-$t$) result $V^{(\!(3)\!)}$,
 \bea
 V^{(\!(3)\!)}&=&F_0(\f,\l,m^2)+F_1(\f,\l,m^2)\,t+F_2(\f,\l,m^2)\,t^2
 +\cdots\;,\label{v3x}   
 \eea
generates, when the running scale $t$ is replaced by the
value given in Eq.~(\ref{t}), all coefficients \{$G_3^{(3)},
G_3^{(4)}, G_3^{(5)}, \cdots$\} in vertical sum for
$V^{(\!(3)\!)}$ in Fig.~1:
 \bea
 \h^3V^{(\!(3)\!)}&=&{\h^3\l^3\ov (4\p)^6}G_3^{(3)}
 +{\h^4\l^4\ov (4\p)^8}G_3^{(4)}y+{\h^5\l^5\ov (4\p)^{10}}G_3^{(5)}y^2
 +\cdots\;.\label{v3y}
 \eea
Each contribution to the right-hand side of Eq.~(\ref{v3y}) is the
next-next-next-to-leading logarithm portion in each loop order
$V^{(l)}$ ($l\ge 3$). 
 
As remarked earlier, with the $L$-loop effective potential and the $(L+1)$-loop
RG functions, one can obtain an RG-improved effective potential which is
exact up to $L$th-to-leading logarithm order \cite{bkmn}. Let us recall
that all the coefficients of $V^{(4)}$, except $G_4^{(4)}$
in Fig.~1, can be known from the $t$ expansions of
$V^{(\!(3)\!)}$, $V^{(\!(3)\!)}$, and $V^{(\!(3)\!)}$. 
Thus, if this unknown coefficient is evaluated by any means, 
one can obtain $V^{(\!(4)\!)}$, the fourth-to-leading logarithm resummation
since the five-loop RG functions are given in Refs.~16--18.

Our analytical result for the RG-improved effective potential, Eq.~(\ref{llse}),
can be adapted to various situations,
i.e., to various ranges of the parameters $m^2$, $\l$, and $\m^2$, except
for $\L$. [This $\L$ appears only in $V^{(\!(0)\!)}$ as an overall additive term
which is immaterial in the analysis of the vacuum structure of a flat-spacetime theory.]
In particular, when the mass-squared parameter $m^2$ is negative, it is interesting 
to investigate how severely the value of $\f_{\rm min}$
changes as the value of the coupling constant $\l$ increases within the  
perturbation-still-reliable range. Further, in order to investigate 
whether our result for the effective potential
is the best one or not, it is necessary to compare the various critical values 
obtained from our result with the existing literature values. 
Work in this direction is in progress \cite{pp}.

\section*{Acknowledgment}
This work was supported by a Korea Research Foundation grant 
(KRF-2000-015-DP0066).

\appendix
\section*{}  
Lower order quantities $V^{(\!(0)\!)}$, $V^{(\!(1)\!)}$, 
and $V^{(\!(2)\!)}$ in Eq.~(\ref{llse}), which have been obtained already
in Ref.~8, are quoted here:
 \bea
 V^{(\!(0)\!)}&=&\L+{m^4\ov 2\l}(1-T^{1/3})
 +{m^2\f^2\ov 2}T^{-1/3}+{\l\f^4\ov 24}T^{-1}\;,\nn\\
 V^{(\!(1)\!)}&=&{\l\ov (4\p)^2}\biggl[{m^4\ov\l}\biggl\{-1+{19\ov 54}T^{1/3}
 +T^{-2/3}\biggl({59\ov 216}+{17\ov 54}\ln T+{1\ov 4}
 \ln S\biggr)\biggr\}\nn\\
 &&+m^2\f^2\biggl\{-{4\ov 27}T^{-1/3}+T^{-4/3}\biggl(-{49\ov 216}
 +{17\ov 54}\ln T +{1\ov 4}
 \ln S\biggr)\biggr\}\nn\\
 &&+\l\f^4\biggl\{{1\ov 216}T^{-1}+T^{-2}\biggl(
 -{85\ov 864}+{17\ov 216}\ln T+{1\ov 16}
 \ln S\biggr)\biggr\}\biggr]\;,\nn\\
 V^{(\!(2)\!)}&=&{\l^2\ov (4\p)^4}\biggl[{m^4\ov\l}\biggl\{{23\ov 30}
 +{6\z(3)\ov 5}-T^{1/3}\biggl({2509\ov 11664}+{2\z(3)\ov 3}\biggr)\nn\\
 &&-T^{-2/3}\biggl({7051\ov 11664}+{2\z(3)\ov 3}+{323\ov 1458}\ln T
 +{19\ov 108}\ln S\biggr)+T^{-5/3}\biggl({5189\ov 29160}
 +{2\z(3)\ov 15}\nn\\
 &&-{731\ov 2916}\ln T+{289\ov 1458}\ln^2 T-{2\ov 27}\ln S
 +{17\ov 54}\ln T\ln S+{1\ov 8}\ln^2 S\biggr)\biggr\}
  +m^2\f^2\nn\\
 &&\times\biggl\{T^{-1/3}\biggl({973\ov 11664}+{\z(3)\ov 3}\biggr)
 -T^{-4/3}\biggl({4025\ov 11664}+{2\z(3)\ov 3}+{68\ov 729}\ln T
 +{2\ov 27}\ln S\biggr)\nn\\
 &&+T^{-7/3}\biggl({1475\ov 1458}-{1\ov 2\sqrt{3}}
 {\rm Cl}_2\Bigl({\p\ov 3}\Bigr)+{\z(3)\ov 3}
 -{850\ov 729}\ln T+{289\ov 729}\ln^2 T-{73\ov 108}\ln S\nn\\
 &&+{17\ov 27}\ln T\ln S+{1\ov 4}\ln^2 S\biggr)\biggr\}
 +\l\f^4\biggl\{{5\ov 1728}T^{-1}+T^{-2}\biggl(
 -{197\ov 1728}-{\z(3)\ov 6}\nn\\
 &&+{17\ov 1944}\ln T
 +{1\ov 144}\ln S\biggr)
 +T^{-3}\biggl({131\ov 288}-{1\ov 4\sqrt{3}}
 {\rm Cl}_2\Bigl({\p\ov 3}\Bigr)+{\z(3)\ov 6}
 -{2023\ov 3888}\ln T\nn\\
 &&+{289\ov 1944}\ln^2 T-{23\ov 72}\ln S
 +{17\ov 72}\ln T\ln S+{3\ov 32}\ln^2 S\biggr)\biggr\}\nn\\
 &&+\biggl\{{m^4\ov\l}\biggl({1\ov 4}T^{-2/3}\biggr)
 +m^2\f^2\biggl({1\ov 4}T^{-4/3}\biggr)
 +\l\f^4\biggl({1\ov 16}T^{-2}\biggr)\biggr\}{Z\ov W}\biggr]\;,
 \label{rslt}
 \eea
where the quantities $T$, $W$, $S$, and $Z$ appear in 
Eq.~(\ref{twszp}).

\newpage
%%%%%%%%%%%%%%%%%%%%%%%%%%%%%%%%%%%%%%%%%%%%
% FIGURE                                   %
%%%%%%%%%%%%%%%%%%%%%%%%%%%%%%%%%%%%%%%%%%%%
\vspace{5mm}
\begin{figure} 
\centerline{\epsfig{figure=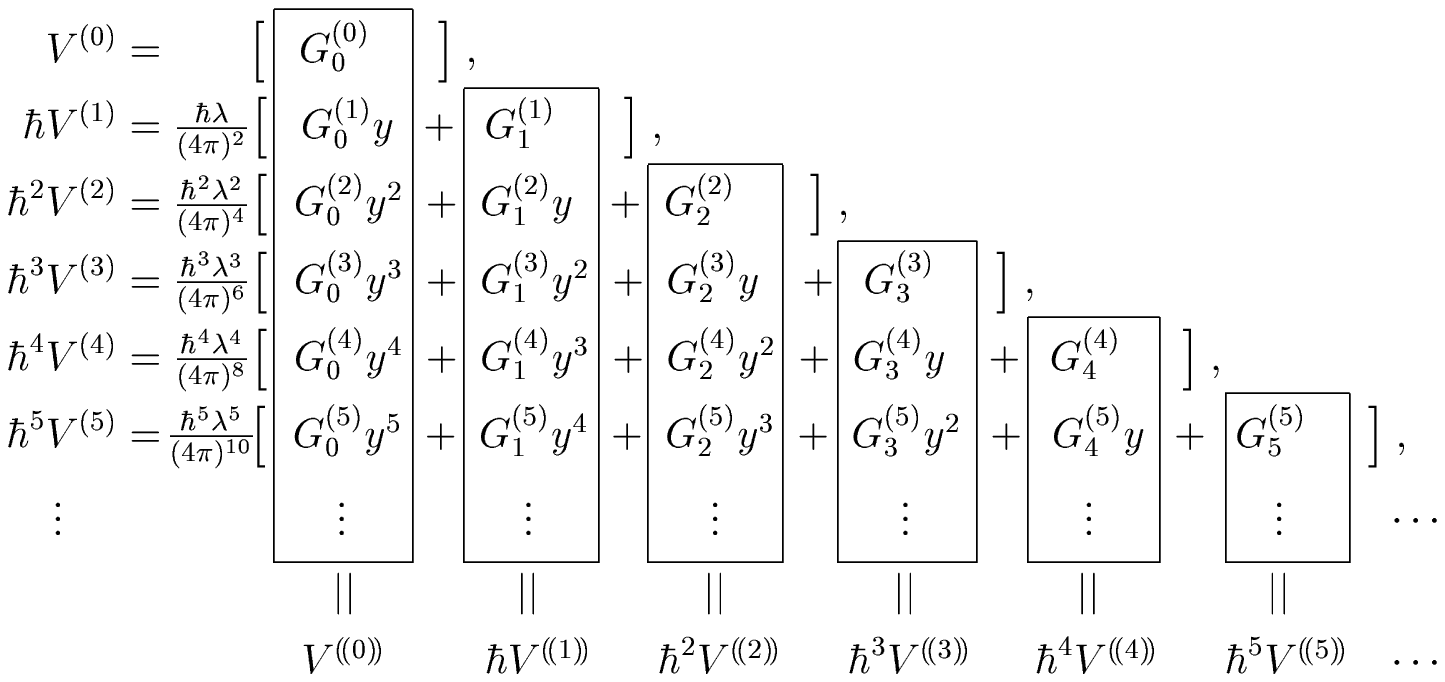,angle=0,width=110mm}}
\vspace{2mm}
\caption{The loop expansion vs. the leading-logarithm  expansion.}

\end{figure}

\end{document}